\documentclass[aps,amssymb,amsmath,
twocolumn,showpacs,superscriptaddress,
groupedaddress,ifmbe]{revtex4-1}

\usepackage{graphicx}
\usepackage{dcolumn}
\usepackage{bm}
\usepackage[table,xcdraw]{xcolor}
\usepackage{nth}
\usepackage[ruled,vlined]{algorithm2e}
\usepackage{framed}
\usepackage{listings}
\usepackage[titletoc,title]{appendix}
\usepackage{tikz}
\usepackage{lipsum}
\usepackage{hyperref}

\hypersetup{
  colorlinks,
  citecolor=red,
  linkcolor=blue,
  urlcolor=blue}

\begin{document}
\title{On the evaluation of the electron repulsion integrals}
\author{A. Ba{\u g}c{\i}}
\email{abagci@pau.edu.tr}
\affiliation{
Instituto de Modelado e Innovaci{\'o}n Tecnol{\'o}gica (IMIT), Facultad de Ciencias Exactas,
Naturales y Agrimensura, Universidad Nacional del Nordeste, Avda. Libertad 5460, W3404AAS, Corrientes, Argentina
\\
Computational Physics Laboratory, Department of Physics, Faculty of Sciences, Pamukkale University, {\c C}amlaraltı, K{\i}n{\i}kl{\i} Campus, Denizli, Turkey}

\author{Gustavo A. Aucar}
\affiliation{Instituto de Modelado e Innovaci{\'o}n Tecnol{\'o}gica (IMIT), Facultad de Ciencias Exactas,
Naturales y Agrimensura, Universidad Nacional del Nordeste, Avda. Libertad 5460, W3404AAS, Corrientes, Argentina}
\begin{abstract}
The electron repulsion integrals over the Slater$-$type orbitals with non$-$integer principal quantum numbers are considered. These integrals are useful in both  non$-$relativistic and relativistic calculations of many$-$electron systems. They involve hyper$-$geometric functions. Due to the non$-$trivial structure of infinite series that are used to define them the hyper$-$geometric functions are practically difficult to compute. Convergence of their series are strictly depends on the values of parameters. Computational issues such as cancellation or round$-$off error emerge. Relationships free from hyper$-$geometric functions for expectation values of Coulomb potential $\left(r_{21}^{-1}\right)$ are derived. These relationships are new and show that the complication coming from two$-$range nature of Laplace expansion for the Coulomb potential is removed. These integrals also form an initial condition for expectation values of a potential with arbitrary power. The electron repulsion integrals are expressed by finite series of power functions. The methodology given here for evaluation of electron repulsion integrals are adapted to multi$-$center integrals.
\begin{description}
\item[Keywords]
Non$-$integer Slater$-$type orbitals, Laplace expansion, Coulomb potential
\item[PACS numbers]
... .
\end{description}
\end{abstract}
\maketitle

\section{Introduction}\label{intro}
Analytical relationships for the electron repulsion integrals are derived using a series representation for the Coulomb interaction via the Laplace expansion and spherical harmonics addition theorem. Finally, the angular parts are separated. They are represented by Clebsch$-$Gordan coefficients. The problem is reduced to a solution of two$-$dimensional radial integral. If Slater$-$type orbitals with non$-$integer principal quantum numbers are considered as a basis then, the radial integrals are expressed in terms of incomplete gamma functions \cite{1_Temme_1996, 2_Bateman_1953} or Gauss's hyper$-$geometric functions \cite{2_Bateman_1953}. For both cases, incomplete gamma or hyper$-$geometric functions the convergence should be investigated in well$-$defined domain. Any representation (recurrence relationship, continued fraction, series expansion formula or asymptotic method) of these special functions depending on the parameters constructs a different domain of convergence \cite{3_Gill_2012, 4_Bujanda_2017, 5_Ansari_2019, 6_Reynolds_2021, 7_Pearson_2017, 8_Fejzullahu_2021}. Besides, the relativistic electrostatic integrals are expressed in terms of its non$-$relativistic counterpart. Each term in the expression comprises two coupled hyper$-$geometric functions that are different from another one. Calculations through the recurrence relationship method for hyper$-$geometric functions results in the emerge of numerous hyper$-$geometric functions with different parameters. This is obviously a complication that need to be fixed.

The present paper recomposes the evaluation of electron repulsion integrals and free the resulting formulas form special functions or any infinite series representation. Another meaning of such procedure is new relationships for the special functions arising in the atomic calculations. This work is organized as follows, In the Section \ref{Revisiting} the procedure for the evaluation of electron repulsion integrals is revisited. The next Section \ref{New} starts with evaluation of radial integrals. The derived new relationships and their mathematical nature are given in this section. The way of solution for the complication mentioned above is given in this section. The advantages of using the formulas presented in this study in atomic and molecular calculations are discussed in the Sections \ref{emdp} and \ref{discuss}.

The Schr{\"o}dinger$-$like differential equation solution has recently been generalized to non$-$integer values of quantum numbers by one of the author \cite{9_Bagci_2022}. The solution is characterized depending to pre$-$determined sequence of quantum numbers. The Slater$-$type orbitals take form according to this solution. There exist four variants of the Slater$-$type orbitals with non$-$integer principal quantum numbers (NSTOs). They are obtained by simplification of the Laguerre polynomials to highest power of $r$. In this work, the most common variant of the NSTOs used in non$-$relativistic \cite{10_Koga_1997, 11_Koga_1997, 12_Koga_1998, 13_Koga_2000, 14_Coskun_2022, 15_Sahin_2022} and relativistic \cite{16_Grant_2000, 17_Grant_2007, 18_Grant_2022} atomic calculations is considered. This variant of NSTOs is a special case of solution for the Schr{\"o}dinger$-$like differential equation while principal quantum numbers $n$, $n \in \mathbb{R}$, $n > 0$ and $\left\{l,m\right\} \in \mathbb{N}$. $\left\{l,m\right\}$ are orbital and magnetic quantum numbers with $0\leq l \leq \lfloor n \rfloor -1$, $-l \leq m \leq l$ and $\lfloor n \rfloor$ represents integer part of $n$, respectively. The relationships obtained here for this type of NSTOs are available to be used for any other variant that is classified in \cite{9_Bagci_2022}.\\
The considered variant of NSTOs are given as,
\begin{multline}\label{eq:1}
\chi_{nlm}\left(\vec{r},\zeta\right)
=R_{n}\left(\zeta,r\right)S_{lm}\left(\theta,\varphi\right)
\\
=N_{n}\left(\zeta\right)r^{n-1}e^{-\zeta r}S_{lm}\left(\theta,\varphi\right),
\end{multline}
here, $\zeta$ is the orbital parameter, $\zeta >0$,
\begin{align}\label{eq:2}
N_{n}\left(\zeta\right)=\dfrac{\left(2\zeta\right)^{n+1/2}}{\sqrt{\Gamma\left(2n+1\right)}},
\end{align}
and the functions $S_{lm}$ are normalized complex $\left(S_{lm} \equiv Y_{lm}, Y^{*}_{lm}=Y_{l-m}\right)$ or real spherical harmonics \cite{19_Condon_1935}.
\begin{figure}[t!]
\includegraphics[width=0.48\textwidth,height=0.25\textheight]{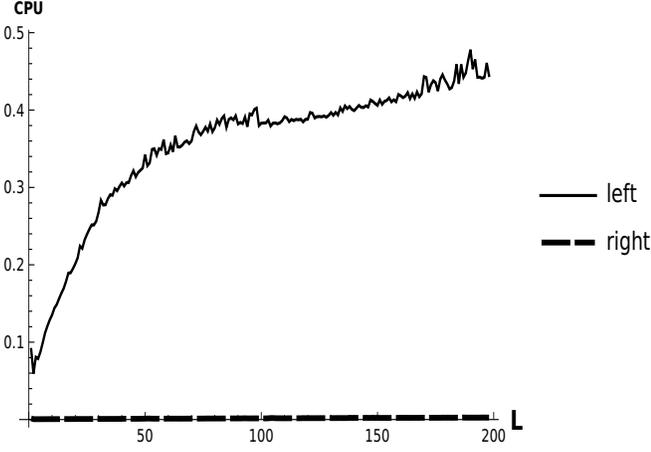}
\caption{Performance of left and right$-$hand sides of the Eq. (\ref{eq:34}). The computation time (CPU) in seconds are plotted versus to $L$, where $n=99.5$, $n^{\prime}=99.51$, $\zeta=1.1$, $\zeta^{\prime}=1.2$. Plot legends $left$, $right$ represent the left and right$-$hand sides of the Eq. (\ref{eq:34}).}
\label{fig:Hypercpu}
\end{figure}
\section{Revisiting the electron repulsion integrals}\label{Revisiting}
One$-$center electron repulsion integrals over STOs are given as \cite{20_Roothaan_1963},
\begin{multline}\label{eq:3}
J_{p_{1},p^{{\prime}}_{1},p_{2},p^{\prime}_{2}}\left(\zeta^{}_{1},\zeta^{\prime}_{1},\zeta^{}_{2},\zeta^{\prime}_{2}\right)
\\
=\int
\chi^{*}_{p_{1}}\left(\bold{r}_{1}\right)
\chi_{p^{\prime}_{1}}\left(\bold{r}_{1}\right)
{\dfrac{1}{r_{12}}}
\chi^{}_{p_{2}}\left(\bold{r}_{2}\right)
\chi^{*}_{p^{\prime}_{2}}\left(\bold{r}_{2}\right)dV_{1}dV_{2},
\end{multline}
with
\begin{align}\label{eq:4}
\chi_{p_{i}}\left(\bold{r}_{i}\right)
=\chi_{n_{i}l_{i}m_{i}}\left(\zeta_{i},\vec{r}_{i}\right).
\end{align}
\begin{align}\label{eq:5}
\chi_{p^{\prime}_{i}}\left(\bold{r}_{i}\right)
=\chi_{n^{\prime}_{i}l^{\prime}_{i}m^{\prime}_{i}}\left(\zeta^{\prime}_{i},\vec{r}_{i}\right).
\end{align}
Taking into account the expansion for the spherical harmonics with the same center,
\begin{multline}\label{eq:6}
R_{n}\left(\zeta,r_{1}\right)S_{lm}\left(\theta_{1},\varphi_{1}\right)
R_{n^{\prime}}\left(\zeta^{\prime},r_{1}\right)S^{*}_{l^{\prime}m^{\prime}}\left(\theta_{1},\varphi_{1}\right)
\\
=R_{n}\left(\zeta,r\right)R_{n^{\prime}}\left(\zeta,r\right)
\sum_{LM}\sqrt{\left(\dfrac{2L+1}{4\pi}\right)}C^{L\vert M \vert}\left(lm,l^{\prime}m^{\prime}\right)
\\
\times A^{M}_{mm^{\prime}}S^{*}_{LM}\left(\theta_{1},\varphi_{1}\right)
\end{multline}
and Laplace expansion for Coulomb interaction,
\begin{align}\label{eq:7}
\dfrac{1}{r_{12}}=\sum_{LM}\left(\dfrac{4\pi}{2L+1}\right)\dfrac{r_{<}^{L}}{r_{>}^{L+1}}S_{LM}\left(\theta_{1},\varphi_{1}\right)S^{*}_{LM}\left(\theta_{2},\varphi_{2}\right),
\end{align}
the electron repulsion integrals finally are expressed as follows,
\begin{multline}\label{eq:8}
J_{n_{1}l_{1}m_{1},n^{{\prime}}_{1}l^{{\prime}}_{1}m^{{\prime}}_{1},n_{2}l_{2}m_{2},n^{\prime}_{2}l^{\prime}_{2}m^{\prime}_{2}}\left(\zeta^{}_{1},\zeta^{\prime}_{1},\zeta^{}_{2},\zeta^{\prime}_{2}\right)
\\
=\sum_{LM}
C^{L\vert M \vert}\left(l_{1}m_{1},l^{\prime}_{1}m^{\prime}_{1}\right)
C^{L\vert M \vert}\left(l_{2}m_{2},l^{\prime}_{2}m^{\prime}_{2}\right)
\\
\times A^{M}_{m_{1}m^{\prime}_{1}}A^{M}_{m_{2}m^{\prime}_{2}}
R^{L}_{n_{1},n^{\prime}_{1},n_{2},n^{\prime}_{2}}\left(\zeta^{}_{1},\zeta^{\prime}_{1},\zeta^{}_{2},\zeta^{\prime}_{2}\right).
\end{multline}
Here, 
\begin{align*}
Max\left(\vert l_{1}-l^{\prime}_{1} \vert, \vert l_{2}-l^{\prime}_{2}\vert \right) \leq L \leq Min\left(l_{1}+l^{\prime}_{1},l_{2}+l^{\prime}_{2}\right),
\end{align*}
\begin{align*}
-L \leq M \leq L.
\end{align*}
Please see \cite{21_Guseinov_1970, 22_Guseinov_1995}, for Gaunt $C^{L\vert M \vert}$ and $A^{M}_{mm'{\prime}}$ coefficients that arise from product of two spherical harmonics. The remaining radial integrals $R^{L}$ are defined as \cite{23_Pitzer_1982},
\begin{multline}\label{eq:9}
R^{L}_{n_{1},n^{\prime}_{1},n_{2},n^{\prime}_{2}}\left(\zeta^{}_{1},\zeta^{\prime}_{1},\zeta^{}_{2},\zeta^{\prime}_{2}\right)
\\
=\int_{0}^{\infty}\int_{0}^{\infty}
r_{1}^{n_{1}+n^{\prime}_{1}}e^{-\left(\zeta^{}_{1}+\zeta^{\prime}_{1}\right){r_{1}}}
\left(\dfrac{r_{<}^{L}}{r_{>}^{L+1}}\right)
\\
\times r_{2}^{n_{2}+n^{\prime}_{2}}e^{-\left(\zeta^{}_{2}+\zeta^{\prime}_{2}\right){r_{2}}}
dr_{1}dr_{2}.
\end{multline}
So far obtained analytical expressions for the radial integrals given in the Eq. (\ref{eq:9}) were in terms of incomplete gamma functions or Gauss's hyper$-$geometric functions. The method of solution for the Eq. (\ref{eq:9}) is detailed here as it is important for our discussion in the next section on how it can be used for molecular integrals.\\
Explicit form of the Eq. (\ref{eq:9}) is written as,
\begin{multline}\label{eq:10}
R^{L}_{n_{1},n^{\prime}_{1},n_{2},n^{\prime}_{2}}\left(\zeta^{}_{1},\zeta^{\prime}_{1},\zeta^{}_{2},\zeta^{\prime}_{2}\right)
\\
=\int_{0}^{\infty}\Bigg\{ \int_{0}^{r_{2}} r_{1}^{n_{1}+n^{\prime}_{1}}e^{-\left(\zeta^{}_{1}+\zeta^{\prime}_{1}\right)r_{1}}\dfrac{r_{1}^{L}}{r_{2}^{L+1}}dr_{1}
\\
+\int_{r_{2}}^{\infty}r_{1}^{n_{1}+n^{\prime}_{1}}e^{-\left(\zeta^{}_{1}+\zeta^{\prime}_{1}\right)r_{1}}\dfrac{r_{2}^{L}}{r_{1}^{L+1}}dr_{1}
\Bigg\}
\\
\times r_{2}^{n_{2}+n^{\prime}_{2}}e^{-\left(\zeta^{}_{2}+\zeta^{\prime}_{2}\right)r_{2}}dr_{2}.
\end{multline}
By applying rule of definite integral to the first term and changing the variable as $x=r_{1}/r_{2}$, the following expression is obtained,
\begin{multline}\label{eq:11}
R^{L}_{n_{1},n^{\prime}_{1},n_{2},n^{\prime}_{2}}\left(\zeta^{}_{1},\zeta^{\prime}_{1},\zeta^{}_{2},\zeta^{\prime}_{2}\right)
\\
=\dfrac{\Gamma\left(n_{1}+n^{\prime}_{1}+L+1\right)}{\left(\zeta^{}_{1}+\zeta^{\prime}_{1}\right)^{n_{1}+n^{\prime}_{1}+L+1}}
\int_{0}^{\infty} r_{2}^{n_{2}+n^{\prime}_{2}-L-1}e^{-\left(\zeta^{}_{2}+\zeta^{\prime}_{2}\right)r_{2}}dr_{2}
\\
-\int_{0}^{\infty}
\Bigg[\mathcal{A}_{n_{1}+n^{\prime}_{1}+L}\left[\left(\zeta^{}_{1}+\zeta^{\prime}_{1}\right)r_{2} \right]
\\
-\mathcal{A}_{n_{1}+n^{\prime}_{1}-L-1}\left[\left(\zeta^{}_{1}+\zeta^{\prime}_{1}\right)r_{2} \right] \Bigg]
\\
\times r_{2}^{n_{1}+n^{\prime}_{1}+n_{2}+n^{\prime}_{2}}e^{-\left(\zeta_{2}+\zeta^{\prime}_{2}\right)r_{2}}dr_{2}
\end{multline}
where,
\begin{align}\label{eq:12}
\mathcal{A}_{n}\left[p\right]=p^{-n-1}\Gamma\left[n+1,p\right].
\end{align}
This form of integrals are used to represent the electron repulsion integrals in terms of both incomplete gamma functions or hyper$-$geometric functions. Inserting the Eq. (\ref{eq:12}) into the Eq. (\ref{eq:11}), after a slightly manipulation we have,
\begin{multline}\label{eq:13}
R^{L}_{n_{1},n^{\prime}_{1},n_{2},n^{\prime}_{2}}\left(\zeta^{}_{1},\zeta^{\prime}_{1},\zeta^{}_{2},\zeta^{\prime}_{2}\right)
=\Gamma\left(n_{1}+n^{\prime}_{1}+L+1\right)
\\
\int_{0}^{\infty} \Bigg[
\dfrac{1}{\left(\zeta^{}_{1}+\zeta^{\prime}_{1}\right)^{n_{1}+n^{\prime}_{1}+L+1}}
\dfrac{\gamma\left[n_{1}+n^{\prime}_{1}+L+1,\left(\zeta^{}_{1}+\zeta^{\prime}_{1}\right)r_{2}\right]}{\Gamma\left(n_{1}+n^{\prime}_{1}+L+1\right)}
\\
+\dfrac{1}{\left(\zeta^{}_{1}+\zeta^{\prime}_{1}\right)^{n_{1}+n^{\prime}_{1}-L}}
\dfrac{\Gamma\left[n_{1}+n^{\prime}_{1}-L,\left(\zeta^{}_{1}+\zeta^{\prime}_{1}\right)r_{2}\right]}{\Gamma\left(n_{1}+n^{\prime}_{1}+L+1\right)}r_{2}^{2L+1} \Bigg]
\\
\times r_{2}^{n_{2}+n^{\prime}_{2}-L-1}e^{-\left(\zeta^{}_{2}+\zeta^{\prime}_{2}\right)r_{2}}dr_{2}.
\end{multline}
$\Gamma\left[n,x\right]$, $\gamma\left[n,x\right]$ are the upper and lower incomplete gamma functions. Note that the first and second terms in the Eq. (\ref{eq:13}) represents the integrals given the Eq. (\ref{eq:10}) over the ranges $r_{1} \in \left[0,r_{2}\right]$, $r_{1} \in \left[r_{2}, \infty\right]$, respectively.
\begin{widetext}
By using the following relationships between incomplete gamma functions and hyper$-$geometric functions \cite{24_Gradshteyn_2007};
\begin{align}\label{eq:14}
\dfrac{1}{a^{n}}\int_{0}^{\infty}x^{m-1}e^{-bx}\Gamma\left[n,ax\right]dx
=\dfrac{\Gamma\left(m+n\right)}{m\left(a+b\right)^{m+n}}
{_2}F_{1}\left[1,m+n,m+1,\frac{b}{a+b}\right],
\end{align}
\begin{align}\label{eq:15}
\dfrac{1}{a^{n}}\int_{0}^{\infty}x^{m-1}e^{-bx}\gamma\left[n,ax\right]dx
=\dfrac{\Gamma\left(m+n\right)}{n\left(a+b\right)^{m+n}}
{_2}F_{1}\left[1,m+n,n+1,\frac{a}{a+b}\right],
\end{align}
The electron repulsion integrals take the form that \cite{25_Allouche_1974},
\begin{multline}\label{eq:16}
R^{L}_{n_{1},n^{\prime}_{1},n_{2},n^{\prime}_{2}}\left(\zeta^{}_{1},\zeta^{\prime}_{1},\zeta^{}_{2},\zeta^{\prime}_{2}\right)
=\dfrac{\Gamma\left(n_{1}+n^{\prime}_{1}+n_{2}+n^{\prime}_{2}+1\right)}{\left(\zeta_{1}+\zeta^{\prime}_{1}+\zeta^{}_{2}+\zeta^{\prime}_{2}\right)^{n_{1}+n^{\prime}_{1}+n_{2}+n^{\prime}_{2}+1}}
\Bigg\{
\dfrac{1}{n_{1}+n^{\prime}_{1}+L+1}
\\
\times {_2}F_{1}\left[1,n_{1}+n^{\prime}_{1}+n_{2}+n^{\prime}_{2}+1,n_{1}+n^{\prime}_{1}+L+2,\frac{\zeta^{}_{1}+\zeta^{\prime}_{1}}{\zeta^{}_{1}+\zeta^{\prime}_{1}+\zeta^{}_{2}+\zeta^{\prime}_{2}}\right]
+\dfrac{1}{n_{2}+n^{\prime}_{2}+L+1}
\\
\times {_2}F_{1}\left[1,n_{1}+n^{\prime}_{1}+n_{2}+n^{\prime}_{2}+1,n_{2}+n^{\prime}_{2}+L+2,\frac{\zeta^{}_{2}+\zeta^{\prime}_{2}}{\zeta^{}_{1}+\zeta^{\prime}_{1}+\zeta^{}_{2}+\zeta^{\prime}_{2}}\right]
\Bigg\}.
\end{multline}
\end{widetext}
The Eq. (\ref{eq:16}) was used to perform Roothaan$-$Hartree$-$Fock calculations \cite{10_Koga_1997, 11_Koga_1997, 12_Koga_1998, 13_Koga_2000, 14_Coskun_2022, 15_Sahin_2022} via Mathematica programming language \cite{14_Coskun_2022, 15_Sahin_2022} and modified version of the Pitzer's program \cite{10_Koga_1997, 11_Koga_1997, 12_Koga_1998, 13_Koga_2000}. We know that the Mathematica programming language takes into account every special condition in computation for hyper$-$geometric functions. In the unpublished modified version of Pitzer's program, the computation for hyper$-$geometric functions using the Eq. (\ref{eq:16}) on the other hand, were not detailed. A computer program package \cite{26_Johansson_2019} has recently published for computation of hyper$-$geometric functions. Although the method of computation is complicated, it has been well studied in \cite{26_Johansson_2019}. In the present paper necessity for calculation for hyper$-$geometric functions is eliminated.
\pagebreak
\section{New relationships for the electron repulsion integrals}\label{New}
By replacing the $n_{1}+n^{\prime}_{1}$ with $n$, $n_{2}+n^{\prime}_{2}$ with $n^{\prime}$ and $\zeta_{1}+\zeta^{\prime}_{1}$ with $\zeta$, $\zeta_{2}+\zeta^{\prime}_{2}$ with $\zeta^{\prime}$, the Eq. (\ref{eq:16}) has a simpler form as,
\begin{multline}\label{eq:17}
R^{L}_{n,n^{\prime}}\left(\zeta,\zeta^{\prime}\right)
=\dfrac{\Gamma\left(n+n^{\prime}+1\right)}{\left(\zeta+\zeta^{\prime}\right)^{n+n^{\prime}+1}}
\Bigg\{
\dfrac{1}{n+L+1}
\\
\times {_2}F_{1}\left[1,n+n^{\prime}+1,n+L+2;\frac{\zeta}{\zeta+\zeta^{\prime}}\right]
+\dfrac{1}{n^{\prime}+L+1}
\\
\times {_2}F_{1}\left[1,n+n^{\prime}+1,n^{\prime}+L+2;\frac{\zeta^{\prime}}{\zeta+\zeta^{\prime}}\right]
\Bigg\}.
\end{multline}
We start by using the following representation of the hyper$-$geometric functions \cite{2_Bateman_1953},
\begin{multline}\label{eq:18}
{_2}F_{1}\left[1,b,c;z \right]
=\dfrac{2\Gamma\left(c\right)}{\Gamma\left(b\right)\Gamma\left(c-b\right)}
\\
\times \int_{0}^{\pi/2}
\dfrac{\left(sin t\right)^{2b-1}\left(cos t\right)^{2c-2b-1}}{\left(1-zsin^{2}t\right)}dt.
\end{multline}
Solution for the integral given in the Eq. (\ref{eq:18}) is obtained as follows,
\begin{multline}\label{eq:19}
{_2}F_{1}\left[1,b,c;z \right]
=\dfrac{\Gamma\left(c\right)}{\Gamma\left(b\right)\Gamma\left(c-b\right)}\dfrac{\pi}{sin\left[\pi\left(b-c\right)\right]}
\\
\Bigg[-\left(1-z\right)^{c-b-1}z^{1-c}+\dfrac{\Gamma\left(b\right)}{\Gamma\left(c-1\right)}
\\
\times \dfrac{1}{\Gamma\left(b-c+2\right)}
{_2}F_{1}\left[1,b,b-c+2;z \right]
\Bigg].
\end{multline}
For the hyper$-$geometric functions arising in the Eq. (\ref{eq:17}) we obtain,
\begin{multline}\label{eq:20}
F_{nn^{\prime}}^{L}\left(\zeta,\zeta^{\prime}\right)
={_2}F_{1}\left[1,n+n^{\prime}+1,n+L+2;\frac{\zeta}{\zeta+\zeta^{\prime}}\right]
\\
={^{1}}f_{nn^{\prime}}^{L}\hspace{1mm}{_2}F_{1}\left[1,n+n^{\prime}+1,n^{\prime}-L+1;\frac{\zeta^{\prime}}{\zeta+\zeta^{\prime}}\right]\\
-g_{nn^{\prime}}^{L}\left(\zeta,\zeta^{\prime}\right),
\end{multline}
Here,
\begin{align}\label{eq:21}
{^{1}}f_{nn^{\prime}}^{L}=
\dfrac{\pi csc\left[\left(-n^{\prime}+L\right)\pi\right]}{\Gamma\left(-n^{\prime}+L+1\right)}\dfrac{\left(n+L+1\right)}{\Gamma\left(n^{\prime}-L+1\right)},
\end{align}
\begin{multline}\label{eq:22}
g_{nn^{\prime}}^{L}\left(\zeta,\zeta^{\prime}\right)
=f_{nn^{\prime}}\dfrac{\Gamma\left(n^{\prime}-L+1\right)\Gamma\left(n+L+1\right)}{\Gamma\left(n+n^{\prime}+1\right)}
\\
\times \left(\frac{\zeta}{\zeta+\zeta^{\prime}}\right)^{-n-L-1}
\left(\frac{\zeta^{\prime}}{\zeta+\zeta^{\prime}}\right)^{-n^{\prime}+L}.
\end{multline}
If the second term in the Eq. (\ref{eq:17}) is considered then,
${^{1}}f_{nn^{\prime}}^{L}\rightarrow {^{1}}f_{n^{\prime}n}^{L}$, $g_{nn^{\prime}}^{L}\left(\zeta, \zeta^{\prime}\right) \rightarrow g_{n^{\prime}n}^{L}\left(\zeta^{\prime}, \zeta\right)$. Two types of expressions for electron repulsion integrals are derived from the Eq. (\ref{eq:20}). The recurrence relationships given below for each expression leads to either increasing $n-L+1$ and $n^{\prime}-L+1$ to $n+L+2$ and $n^{\prime}+L+2$, or decreasing $n+L+2$ and $n^{\prime}+L+2$ to $n-L+1$ and $n^{\prime}-L+1$, respectively.
\begin{multline}\label{eq:23}
{_2}F_{1}\left[1,b,c;z\right]
=\dfrac{\left(1-c\right)_{m}}{\left(b-c+1\right)_{m}}\left(\dfrac{z-1}{z}\right)^{m}
\\
\times {_2}F_{1}\left[1,b,c-m;z\right]
\\
+\left(\dfrac{1}{z}\right)
\sum_{k=1}^{m}\dfrac{\left(1-c\right)_{k}}{\left(b-c+1\right)_{k}}\left(\dfrac{z-1}{z}\right)^{k-1}
\end{multline}
\begin{multline}\label{eq:24}
{_2}F_{1}\left[1,b,c-m;z\right]
=\dfrac{\left(b-c+1\right)_{m}}{\left(1-c\right)_{m}}\left(\dfrac{z}{z-1}\right)^{m}
\\
\times {_2}F_{1}\left[1,b,c;z\right]
-\left(\dfrac{z}{z-1}\right)^{m}\left(\dfrac{1}{z}\right)
\\
\times \sum_{k=1}^{m}\dfrac{\left(b-c+1+k\right)_{m-k}}{\left(1-c+k\right)_{m-k}}\left(\dfrac{z-1}{z}\right)^{k-1}.
\end{multline}
Finally, we obtain four different types of expressions for electron repulsion integrals. Two of them for hyper$-$geometric functions involve $n+L+2$, $n^{\prime}+L+2$ and other two of them for $n-L+1$, $n^{\prime}-L+1$:\\
For electron repulsion integrals with hyper$-$geometric functions involve $n+L+2$ and $n^{\prime}+L+2$,
\begin{multline}\label{eq:25}
R^{L}_{n,n^{\prime}}\left(\zeta,\zeta^{\prime}\right)
=\dfrac{\Gamma\left(n+n^{\prime}+1\right)}{\left(\zeta+\zeta^{\prime}\right)^{n+n^{\prime}+1}}\dfrac{1}{n^{\prime}+L+1}
\\
\times {_2}F_{1}\left[1,n+n^{\prime}+1,n^{\prime}+L+2;\frac{\zeta^{\prime}}{\zeta+\zeta^{\prime}}\right]{^{1}}h_{nn^{\prime}}^{L}\left(\zeta,\zeta^{\prime}\right)
\\
-\dfrac{\Gamma\left(n+n^{\prime}+1\right)}{\left(\zeta+\zeta^{\prime}\right)^{n+n^{\prime}+1}}\dfrac{1}{n+L+1}
\\
\times \Bigg\{g_{nn^{\prime}}^{L}\left(\zeta,\zeta^{\prime}\right)+{^{1}}l_{nn^{\prime}}^{L}\left(\zeta,\zeta^{\prime}\right)
\Bigg\},
\end{multline}
\begin{multline}\label{eq:26}
R^{L}_{n,n^{\prime}}\left(\zeta,\zeta^{\prime}\right)
=\dfrac{\Gamma\left(n+n^{\prime}+1\right)}{\left(\zeta+\zeta^{\prime}\right)^{n+n^{\prime}+1}}\dfrac{1}{n+L+1}
\\
\times {_2}F_{1}\left[1,n+n^{\prime}+1,n+L+2;\frac{\zeta}{\zeta+\zeta^{\prime}}\right]{^{1}}h_{n^{\prime}n}^{L}\left(\zeta^{\prime},\zeta\right)
\\
-\dfrac{\Gamma\left(n+n^{\prime}+1\right)}{\left(\zeta+\zeta^{\prime}\right)^{n+n^{\prime}+1}}\dfrac{1}{n^{\prime}+L+1}
\\
\times \Bigg\{g_{n^{\prime}n}^{L}\left(\zeta^{\prime},\zeta\right)+{^{1}}l_{n^{\prime}n}^{L}\left(\zeta^{\prime},\zeta\right)
\Bigg\},
\end{multline}
Notice that,
$R^{L}_{n,n^{\prime}}\left(\zeta,\zeta^{\prime}\right)=R^{L}_{n^{\prime},n}\left(\zeta^{\prime},\zeta\right)$.\\
For electron repulsion integrals with hyper$-$geometric functions involve $n-L+1$, $n^{\prime}-L+1$,
\begin{multline}\label{eq:27}
R^{L}_{n,n^{\prime}}\left(\zeta,\zeta^{\prime}\right)
=\dfrac{\Gamma\left(n+n^{\prime}+1\right)}{\left(\zeta+\zeta^{\prime}\right)^{n+n^{\prime}+1}}\dfrac{1}{n^{\prime}+L+1}
\\
\times {_2}F_{1}\left[1,n+n^{\prime}+1,n^{\prime}-L+1;\frac{\zeta^{\prime}}{\zeta+\zeta^{\prime}}\right]
{^{2}}h_{nn^{\prime}}^{L}\left(\zeta, \zeta^{\prime}\right)
\\
-\dfrac{\Gamma\left(n+n^{\prime}+1\right)}{\left(\zeta+\zeta^{\prime}\right)^{n+n^{\prime}+1}}\dfrac{1}{n+L+1}
\\
\Bigg\{
g_{nn^{\prime}}^{L}\left(\zeta,\zeta^{\prime}\right)
-{^{2}}l_{nn^{\prime}}^{L}\left(\zeta, \zeta^{\prime}\right)
\Bigg\}
\end{multline}
\begin{multline}\label{eq:28}
R^{L}_{n,n^{\prime}}\left(\zeta,\zeta^{\prime}\right)
=\dfrac{\Gamma\left(n+n^{\prime}+1\right)}{\left(\zeta+\zeta^{\prime}\right)^{n+n^{\prime}+1}}\dfrac{1}{n+L+1}
\\
\times {_2}F_{1}\left[1,n+n^{\prime}+1,n-L+1;\frac{\zeta}{\zeta+\zeta^{\prime}}\right]
{^{2}}h_{n^{\prime}n}^{L}\left(\zeta^{\prime},\zeta\right)
\\
-\dfrac{\Gamma\left(n+n^{\prime}+1\right)}{\left(\zeta+\zeta^{\prime}\right)^{n+n^{\prime}+1}}\dfrac{1}{n^{\prime}+L+1}
\\
\Bigg\{
g_{n^{\prime}n}^{L}\left(\zeta^{\prime},\zeta\right)
-{^{2}}l_{n^{\prime}n}^{L}\left(\zeta^{\prime},\zeta\right)
\Bigg\}
\end{multline}
and the relationship
$R^{L}_{n,n^{\prime}}\left(\zeta,\zeta^{\prime}\right)=R^{L}_{n^{\prime},n}\left(\zeta^{\prime},\zeta\right)$ holds. The functions given in the Eqs. (\ref{eq:25},\ref{eq:26}) are defined as,
\begin{multline}\label{eq:29}
{^{1}}h_{nn^{\prime}}^{L}\left(\zeta, \zeta^{\prime}\right)
=\dfrac{n^{\prime}+L+1}{n+L+1}\dfrac{\left(n-L\right)_{2L+1}}{\left(-n^{\prime}-L-1\right)_{2L+1}}
\\
\left(-\dfrac{\zeta^{\prime}}{\zeta}\right)^{2L+1}
f_{nn^{\prime}}^{L}+1
\end{multline}
\begin{multline}\label{eq:30}
{^{2}}h_{nn^{\prime}}^{L}\left(\zeta, \zeta^{\prime}\right)
=\dfrac{n^{\prime}+L+1}{n+L+1}f_{nn^{\prime}}
\\
+\dfrac{\left(-n^{\prime}-L-1\right)_{2L+1}}{\left(n-L\right)_{2L+1}}\left(-\frac{\zeta}{\zeta^{\prime}}\right)^{2L+1}
\end{multline}
\begin{multline}\label{eq:31}
{^{1}}l_{nn^{\prime}}^{L}\left(\zeta, \zeta^{\prime}\right)
=\dfrac{\pi csc\left[\left(-n^{\prime}+L\right)\pi\right]}{\Gamma\left(-n^{\prime}+L+1\right)}
\dfrac{\left(n+L+1\right)}{\Gamma\left(n^{\prime}-L+1\right)}
\\
\times \left(-\dfrac{\zeta^{\prime}}{\zeta}\right)^{2L+1}
\left(\dfrac{\zeta+\zeta^{\prime}}{\zeta^{\prime}}\right)
\\
\times \sum_{k=1}^{2L+1}\dfrac{\left(n-L+k\right)_{2L+1-k}}{\left(-n^{\prime}-L-1+k\right)_{2L+1-k}}\left(-\dfrac{\zeta}{\zeta^{\prime}}\right)^{k-1}
\end{multline}
\begin{multline}\label{eq:32}
{^{2}}l_{nn^{\prime}}^{L}\left(\zeta, \zeta^{\prime}\right)
=
\dfrac{n+L+1}{n^{\prime}+L+1}
\\
\times \left(\dfrac{\zeta+\zeta^{\prime}}{\zeta^{\prime}}\right)
\sum_{k=1}^{2L+1}\dfrac{\left(-n^{\prime}-L-1\right)_{k}}{\left(n-L\right)_{k}}\left(-\frac{\zeta}{\zeta^{\prime}}\right)^{k}
\end{multline}
The four relationships obtained for electron repulsion integrals are still involve hyper$-$geometric functions although this time only one. Reducing the number of hyper$-$geometric functions decreases the CPU time. On the other hand, extracting the hyper$-$geometric functions in Eqs. (\ref{eq:25}, \ref{eq:26}) we have,
\begin{multline}\label{eq:33}
{_2}F_{1}\left[1,n+n^{\prime}+1,n+L+2;\frac{\zeta}{\zeta+\zeta^{\prime}}\right]
\\
=\dfrac{R^{L}_{n^{\prime},n}\left(\zeta, \zeta^{\prime}\right)+m^{L}_{n^{\prime}n}\left(\zeta^{\prime},\zeta\right)}{e^{L}_{n^{\prime}n}h_{n^{\prime}n}^{L}\left(\zeta^{\prime},\zeta\right)},
\end{multline}
\begin{multline}\label{eq:34}
{_2}F_{1}\left[1,n+n^{\prime}+1,n^{\prime}+L+2;\frac{\zeta^{\prime}}{\zeta+\zeta^{\prime}}\right]
\\
=\dfrac{R^{L}_{n,n^{\prime}}\left(\zeta^{\prime}, \zeta\right)+m^{L}_{nn^{\prime}}\left(\zeta, \zeta^{\prime}\right)}{e^{L}_{nn^{\prime}}h_{nn^{\prime}}^{L}\left(\zeta, \zeta^{\prime}\right)},
\end{multline}
here, 
\begin{align}\label{eq:35}
e^{L}_{nn^{\prime}}
=\dfrac{\Gamma\left(n+n^{\prime}+1\right)}{\left(\zeta+\zeta^{\prime}\right)^{n+n^{\prime}+1}}\dfrac{1}{n^{\prime}+L+1},
\end{align}
and,
\begin{align}\label{eq:36}
m^{L}_{nn^{\prime}}\left(\zeta, \zeta^{\prime}\right)
=e^{L}_{nn^{\prime}}
\left\{g_{nn^{\prime}}^{L}\left(\zeta,\zeta^{\prime}\right)+l_{nn^{\prime}}^{L}\left(\zeta,\zeta^{\prime}\right)
\right\}
\end{align}
Using the following recurrence relationship,
\begin{multline}\label{eq:37}
{_2}F_{1}\left[1,b,c;z\right]
=\dfrac{\left(c-1\right)\left[2-c\left(a+b-2c+3\right)z\right]}{\left(a-c+1\right)\left(b-c+1\right)z}
\\
\times {_2}F_{1}\left[1,b,c-1;z\right]
+\dfrac{\left(c-1\right)\left(c-2\right)\left(1-z\right)}{\left(a-c+1\right)\left(b-c+1\right)z}
\\
\times {_2}F_{1}\left[1,b,c-2;z\right]
\end{multline}
We obtain,
\begin{multline}\label{eq:38}
\mathfrak{R}^{L+2}_{n,n^{\prime}}\left(\zeta,\zeta^{\prime}\right)
=\dfrac{\left(n+L+3\right)}{\zeta\left(n+L+2\right)\left(-n^{\prime}+L+2\right)}
\\
\times \bigg\{
\zeta^{\prime} \left(n+L+2\right)\mathfrak{R}^{L}_{n,n^{\prime}}\left(\zeta,\zeta^{\prime}\right)
\\
+\left[\zeta \left(-n^{\prime}+L+1\right)-\zeta^{\prime}\left(n+L+2\right)\right]
\\
\times
\mathfrak{R}^{L+1}_{n,n^{\prime}}\left(\zeta,\zeta^{\prime}\right)
\bigg\}.
\end{multline}
\begin{multline}\label{eq:39}
\mathfrak{R}^{L+2}_{n,n^{\prime}}\left(\zeta^{\prime},\zeta\right)
=\dfrac{\left(n^{\prime}+L+3\right)}{\zeta^{\prime}\left(n^{\prime}+L+2\right)\left(-n+L+2\right)}
\\
\times \bigg\{
\zeta \left(n^{\prime}+L+2\right)\mathfrak{R}^{L}_{n,n^{\prime}}\left(\zeta^{\prime},\zeta\right)
\\
+\left[\zeta^{\prime} \left(-n+L+1\right)-\zeta\left(n^{\prime}+L+2\right)\right]
\\
\times
\mathfrak{R}^{L+1}_{n,n^{\prime}}\left(\zeta^{\prime},\zeta\right)
\bigg\},
\end{multline}
where,
\begin{align}\label{eq:40}
\mathfrak{R}^{L}_{n,n^{\prime}}\left(\zeta,\zeta^{\prime}\right)
=\dfrac{R^{L}_{n,n^{\prime}}\left(\zeta,\zeta^{\prime}\right)+m^{L}_{n^{\prime}n}\left(\zeta^{\prime}, \zeta\right)}{e^{L}_{n^{\prime}n}h_{n^{\prime}n}^{L}\left(\zeta^{\prime},\zeta\right)}.
\end{align}
Similar routine is used to derive the formulas for decreasing $L$ through the Eqs. (\ref{eq:27}, \ref{eq:28}).\\
The recurrence relationships given in the Eqs. (\ref{eq:38}, \ref{eq:39}) become drastically fast and much simpler (\textcolor{red}{See the Figure \ref{fig:Hypercpu}}). For $L=0$ we have,
\begin{multline}\label{eq:41}
\mathfrak{R}^{0}_{n,n^{\prime}}\left(\zeta,\zeta^{\prime}\right)=
\dfrac{R^{0}_{n,n^{\prime}}\left(\zeta,\zeta^{\prime}\right)+m^{0}_{nn^{\prime}}\left(\zeta, \zeta^{\prime}\right)}{e^{0}_{nn^{\prime}}h_{nn^{\prime}}^{0}\left(\zeta, \zeta^{\prime}\right)}
\\
=\left(n+1\right)\left(\dfrac{\zeta}{\zeta+\zeta^{\prime}}\right)^{-n-1}
\left(\dfrac{\zeta^{\prime}}{\zeta+\zeta^{\prime}}\right)^{-n^{\prime}}
\\
\Bigg\{
\dfrac{\Gamma\left(n+1\right)\Gamma\left(n^{\prime}\right)}{\Gamma\left(n+n^{\prime}+1\right)}
-B_{n^{\prime}n+1}\left(\dfrac{\zeta^{\prime}}{\zeta+\zeta^{\prime}}\right)
\Bigg\},
\end{multline}
with $B_{nn^{\prime}}$ are the incomplete beta functions. 
\begin{align}\label{eq:42}
B_{nn^{\prime}}\left(z\right)=B_{nn^{\prime}}-B_{n^{\prime}n}\left(1-z\right).
\end{align}
For $L=1$, the Eq. (\ref{eq:23}) is used. The reults obtained from Eqs. (\ref{eq:38}) and Eq. (\ref{eq:39}) are directly used in the Eq. (\ref{eq:17}). Or, one can prefer to obtain the result just by calculating the $m_{nn^{\prime}}$ for $L+2$. All the integrals over $L$ arising in the Eq. (\ref{eq:8}) are calculated once and for all without any infinite summation or hyper$-$geometric functions by using the recurrence relationship for ${^{1}}l^{L}_{nn^{\prime}}$ over $L$ (\textcolor{red}{See the appendix \ref{recl}}).

The relationships derived in this section are specially useful in relativistic calculations. There, the radial integrals for charge density of a relativistic basis function,
\begin{multline}\label{eq:43}
\chi_{nlj}\left(r,\zeta\right)
\chi_{nlj}\left(r,\zeta^{\prime}\right)
\\
=\bigg( A_{nlj}r^{n}+\zeta B_{nlj}r^{n+1}\bigg)
\bigg(A_{n^{\prime}l^{\prime}j^{\prime}}r^{n^{\prime}}+\zeta^{\prime}B_{n^{\prime}l^{\prime}j^{\prime}}r^{n^{\prime}+1}\bigg)
\\
\times e^{-\left(\zeta+\zeta^{\prime}\right) r},
\end{multline}
with $j$ is the total angular momentum quantum numbers, requires not only using the recurrence relationships over $c$ in hyper$-$geometric functions, but also over $b$. Since the radial integrals consist product of two charge densities, recurrence relationships over multiple hyper$-$geometric functions are mandatory. On the other hand, the recurrence relationships for ${^{1}}l^{L}_{nn^{\prime}}$ over $n$, $n^{\prime}$ together with the Eq. (\ref{eq:38}) or the Eq. (\ref{eq:39}) easily solve this complication. More details are given in the following section.
\section{Extending the solution to relativistic atomic calculations}\label{emdp}
In this section we consider the Laplace expansions for arbitrary powers $r^{\mu}$, $\mu=-1,-2,-3,...$ and their applications in solution of the Dirac equation for many$-$electron systems.

The Slater$-$type spinor orbitals (STSOs) the relativistic are considered analogues of Slater$-$type functions with non$-$integer principal quantum numbers. They are defined as \cite{27_Bagci_2016, 28_Bagci_2020},
\begin{align}\label{eq:44}
X_{nljm}\left(\zeta, \vec{r} \right)
=\begin{pmatrix}
\chi_{nljm}^{\beta 0}\left(\zeta, \vec{r}\right)
\vspace{1.5 mm}\\
\chi_{nljm}^{\beta 1}\left(\zeta, \vec{r}\right)
\vspace{1.5 mm}\\
\chi_{nljm}^{-\beta 0}\left(\zeta, \vec{r}\right)
\vspace{1.5 mm}\\
\chi_{nljm}^{-\beta 1}\left(\zeta, \vec{r}\right)
\end{pmatrix},
\end{align}
here:
\begin{align}\label{eq:45}
\chi_{nljm}^{\beta \varepsilon}\left(\zeta, \vec{r}\right)
=f^{\beta}_{nlj}(\zeta,r) \Omega_{ljm}^{\beta \varepsilon} \left(\theta, \vartheta\right),
\end{align}
\begin{align}\label{eq:46}
f^{\beta}_{nlj}(\zeta,r)
=\left\{{A_{nlj}^{\beta}r^{n}+\zeta B_{nlj}^{\beta}r^{n+1}}\right\}e^{-\zeta r},
\end{align}
with, $\beta$ represents large$-$ and small$-$components of STSOs. The $\Omega_{ljm}^{\beta \varepsilon}$ are the spin $\frac{1}{2}$ spinor spherical harmonics \cite{17_Grant_2007}.\\
The radial charge density of STSOs defined as,
\begin{multline}\label{eq:47}
g^{\beta \beta^{\prime}}_{nljm,n^{\prime}l^{\prime}j^{\prime}}\left(\zeta, \zeta^{\prime},r\right)
\\
=\begin{pmatrix}
A^{\beta}_{nlj}A^{\beta^{\prime}}_{n^{\prime}l^{\prime}j^{\prime}}
\vspace{1.5 mm}\\
\zeta^{\prime} A^{\beta}_{nlj}B^{\beta^{\prime}}_{n^{\prime}l^{\prime}j^{\prime}}
\vspace{1.5 mm}\\
\zeta B^{\beta}_{nlj}A^{\beta^{\prime}}_{n^{\prime}l^{\prime}j^{\prime}}
\vspace{1.5 mm}\\
\zeta \zeta^{\prime} B^{\beta}_{nlj}B^{\beta^{\prime}}_{n^{\prime}l^{\prime}j^{\prime}}
\end{pmatrix}^{\dagger}
\begin{pmatrix}
r^{n+n^{\prime}}
\vspace{1.5 mm}\\
r^{n+n^{\prime}+1}
\vspace{1.5 mm}\\
r^{n+n^{\prime}+1}
\vspace{1.5 mm}\\
r^{n+n^{\prime}+2}
\end{pmatrix} e^{-\left(\zeta, \zeta^{\prime}\right)}.
\end{multline}
The electrostatic interaction is given by,
\begin{multline}\label{eq:48}
G_{p_{1},p^{{\prime}}_{1},p_{2},p^{\prime}_{2}}\left(\zeta^{}_{1},\zeta^{\prime}_{1},\zeta^{}_{2},\zeta^{\prime}_{2}\right)
\\
=\int
\left(g^{\beta_{1} \beta_{1}^{\prime}}_{p_{1},p^{\prime}_{1}}\left(\zeta^{}_{1}, \zeta_{1}^{\prime},r\right)\right)^{\dagger}
{\dfrac{1}{r_{12}}}
\hspace{1mm}
g^{\beta_{2} \beta_{2}^{\prime}}_{p_{2},p^{\prime}_{2}}\left(\zeta^{}_{2}, \zeta_{2}^{\prime},r\right).
\end{multline}
The Eq. (\ref{eq:48}) is represented in terms of sixteen non$-$relativistic radial integrals over Slater$-$type functions. Due to the electron repulsion integrals in the Eq. (\ref{eq:16}) consists hyper$-$geometric functions, finding a simple relation that is used to express one of the sixteen terms with other is difficult. The Eqs. (\ref{eq:38}, \ref{eq:39}) run for all these terms. What left to avoid additional calculations is to derive recurrence relationships for $m_{nn^{\prime}}^{L}$, more explicitly since it involves a finite summation for $l_{nn^{\prime}}^{L}$ (\textcolor{red}{See the appendix \ref{recl}}).
\subsection{On the integrals involve Laplace expansion for arbitrary power $r^{\mu}$}\label{emdp1}
The differential equation corresponding to radial part of Laplace expansion for arbitrary power and its solution given as \cite{29_Sack_1964},
\begin{multline}\label{eq:49}
\dfrac{\partial \mathcal{R}_{\mu}^{L}}{\partial r_{1}^{2}}
+\dfrac{2}{r_{1}}\dfrac{\partial \mathcal{R}_{\mu}^{L}}{\partial r_{1}}
-L\left(L+1\right)\dfrac{\mathcal{R}_{\mu}^{L}}{r_{1}^{2}}
\\
=\dfrac{\partial \mathcal{R}_{\mu}^{L}}{\partial r_{2}^{2}}
+\dfrac{2}{r_{2}}\dfrac{\partial \mathcal{R}_{\mu}^{L}}{\partial r_{2}}
-L\left(L+1\right)\dfrac{\mathcal{R}_{\mu}^{L}}{r_{2}^{2}},
\end{multline} 
\begin{multline}\label{eq:50}
\mathcal{R}_{\mu}^{L}\left(r_{1},r_{2}\right)
=\dfrac{\left(-\frac{1}{2}\mu \right)_{L}}{\left(\frac{1}{2}\right)_{L}}
r_{>}^{\mu}\left(\dfrac{r_{<}}{r_>}\right)^{L}
\\
\times {_2}F_{1}\left[a,b,L+\frac{3}{2};\frac{r_{<}^{2}}{r_{>}^{2}} \right],
\end{multline}
here, if $\mu=-1,-3,-5,...$, then $a=-\dfrac{1}{2}-\dfrac{1}{2}\mu$ and $b=L-\frac{1}{2}\mu$. If $\mu=-2,-4,-6,...$, then, $a=L-\frac{1}{2}\mu$ and $b=-\dfrac{1}{2}-\dfrac{1}{2}\mu$. For both cases $a \in \mathbb{N}$ and $b-c \in \mathbb{N}$.\\
The radial integrals for $\mathcal{R}_{\mu}^{L}$,
\begin{multline}\label{eq:51}
{^\mu}R^{L}_{n_{1},n^{\prime}_{1},n_{2},n^{\prime}_{2}}\left(\zeta^{}_{1},\zeta^{\prime}_{1},\zeta_{2},\zeta^{\prime}_{2}\right)
\\
=\int_{0}^{\infty}\int_{0}^{\infty}
r_{1}^{n_{1}+n^{\prime}_{1}}e^{-\left(\zeta^{}_{1}+\zeta^{\prime}_{1}\right){r_{1}}}
\mathcal{R}_{\mu}^{L}\left(r_{1}, r_{2}\right)
\\
\times r_{2}^{n_{2}+n^{\prime}_{2}}e^{-\left(\zeta^{}_{2}+\zeta^{\prime}_{2}\right){r_{2}}}
dr_{1}dr_{2}.
\end{multline}
The recurrence relationships given for $\mathcal{R}_{\mu}^{L}$ \cite{29_Sack_1964} are useful to represent the Eq. (\ref{eq:51}) in terms of standard electron repulsion integrals;
\begin{multline}\label{eq:52}
\left(4+2L+\mu\right)\left(2L-2-\mu\right)\mathcal{R}_{\mu+2}^{L}
+2\left(\mu+2\right)^{2}r_{1}^{2}\mathcal{R}_{\mu}^{L}
\\
+2\left(\mu+2\right)^{2}r_{2}^{2}\mathcal{R}_{\mu}^{L}
=\mu\left(\mu+2\right)r_{1}^{2}\mathcal{R}_{\mu-2}^{L}
\\
-\mu\left(\mu+2\right)2r_{1}r_{2}\mathcal{R}_{\mu-2}^{L}
+\mu\left(\mu+2\right)r_{2}^{2}\mathcal{R}_{\mu-2}^{L},
\end{multline}
therefore,
\begin{multline}\label{eq:53}
\left(4+2L+\mu\right)\left(2L-2-\mu\right)\hspace{1mm}
{^{\mu+2}}R^{L}_{n_{1},n^{\prime}_{1},n_{2},n^{\prime}_{2}}
\\
+2\left(\mu+2\right)^{2}\hspace{1mm}
{^{\mu}}R^{L}_{n_{1}+1,n^{\prime}_{1}+1,n_{2},n^{\prime}_{2}}
\\
+2\left(\mu+2\right)^{2}\hspace{1mm}
{^{\mu}}R^{L}_{n_{1},n^{\prime}_{1},n_{2}+1,n^{\prime}_{2}+1}
\\
=\mu\left(\mu+2\right)\hspace{1mm}{^{\mu-2}}R^{L}_{n_{1}+1,n^{\prime}_{1}+1,n_{2},n^{\prime}_{2}}
\\
-\mu\left(\mu+2\right)\hspace{1mm}{^{\mu-2}}R^{L}_{n_{1}+1,n^{\prime}_{1},n_{2}+1,n^{\prime}_{2}}
\\
+\mu\left(\mu+2\right)\hspace{1mm}{^{\mu-2}}R^{L}_{n_{1},n^{\prime}_{1},n_{2}+1,n^{\prime}_{2}+1}
\end{multline}
where, ${^{-1}}R^{L}_{n_{1},n^{\prime}_{1},n_{2},n^{\prime}_{2}}=R^{L}_{n_{1},n^{\prime}_{1},n_{2},n^{\prime}_{2}}$. In \cite{29_Sack_1964} recurrence relationships over $L$ and $\mu, L$ were also derived. They were considered useful due to existence of hyper$-$geometric functions. Here, the basic electron repulsion integrals are free from special functions or infinite series representation.

The Breit interaction is the lower order correction to electron$-$electron Coulomb interaction \cite{17_Grant_2007}. 
\begin{multline}\label{eq:54}
H_{ij}=H^{G}_{ij}+H^{ret}_{ij}
=-\dfrac{\vec{\alpha}_{i}.\vec{\alpha}_{j}}{r_{ij}}
\\
+\dfrac{1}{2}
\left\{
\dfrac{\vec{\alpha}_{i}.\vec{\alpha}_{j}}{r_{ij}}
-\dfrac{\left(\vec{r}_{ij}.\vec{\alpha}_{i}\right)\left(\vec{r}_{ij}.\vec{\alpha}_{j}\right)}{r_{ij}^{3}}
\right\},
\end{multline}
here, $\alpha_{i}$ are the Dirac matrices. The first term, represents the un$-$retarded interaction between two Dirac currents and it includes spin$-$spin, spin$-$other$-$orbit and spin$-$spin interactions. The second term accounts for retardation effects \cite{30_Grant_2014, 31_Rocco_2016}. A computer program code to calculate the Breit interaction was given in \cite{32_Fischer_1991,33_Fischer_2000} via the tensor form representation. The most of resulting radial integrals were directly expressed by electrostatic interaction ${^{-1}}R^{L}$ by simply adding a correction which is a constant. The integrals come from the spin$-$spin, spin$-$other$-$orbit were given as \cite{32_Fischer_1991},
\begin{multline}\label{eq:55}
N^{L}_{n_{1},n_{1}^{\prime},n_{2},n_{2}^{\prime}}\left(\zeta_{1},\zeta_{1}^{\prime},\zeta_{2},\zeta_{2}^{\prime}\right)
\\
=\int_{0}^{\infty}\int_{0}^{\infty}
r_{1}^{n_{1}+n^{\prime}_{1}}e^{-\left(\zeta^{}_{1}+\zeta^{\prime}_{1}\right){r_{1}}}
\left(\dfrac{r_{<}^{L}}{r_{>}^{L+3}}\right)
\\
\times e\left(r_{1}-r_{2}\right)r_{2}^{n_{2}+n^{\prime}_{2}}e^{-\left(\zeta^{}_{2}+\zeta^{\prime}_{2}\right){r_{2}}}
dr_{1}dr_{2},
\end{multline}
\begin{multline}\label{eq:56}
V^{L}_{n_{1},n_{1}^{\prime},n_{2},n_{2}^{\prime}}\left(\zeta_{1},\zeta_{1}^{\prime},\zeta_{2},\zeta_{2}^{\prime}\right)
\\
=\int_{0}^{\infty}\int_{0}^{\infty}
r_{1}^{n_{1}}e^{-\zeta^{}_{1}{r_{1}}}\left(\dfrac{\partial}{\partial r_{1}} r_{1}^{n^{\prime}_{1}}e^{-\zeta^{\prime}_{1}{r_{1}}}\right)
\left(\dfrac{r_{<}^{L}}{r_{>}^{L+3}}\right)
\\
\times r_{2}^{n_{2}+n^{\prime}_{2}}e^{-\left(\zeta^{}_{2}+\zeta^{\prime}_{2}\right){r_{2}}}
dr_{1}dr_{2},
\end{multline}
where,
\begin{align}\label{eq:57}
e\left(x\right)=\left\{
\begin{array}{ll}
1 \hspace{3mm} x >1
\\
0  \hspace{3mm} x \leq 1.
\end{array}
\right.
\end{align}
The Eqs. (\ref{eq:55}, \ref{eq:56}) can also be represented in terms of electrostatic interaction via the Eq. (\ref{eq:53}).
\begin{table}
\caption{\label{tab:Hyperlist} First ten elements of a list obtained from the Eq. (\ref{eq:34}) for hyper$-$geometric functions and used to compare the calculation times in the Figure \ref{fig:Hypercpu}, where $n=99.5$, $n^{\prime}=99.51$, $\zeta=1.1$, $\zeta^{\prime}=1.2$.}
\begin{ruledtabular}
\begin{tabular}{cc}
$L$ & Eq. (\ref{eq:34})
\\
0 & 29.214103839897745 \\
1 & 25.622699228991727 \\
2 & 22.679215280002873 \\
3 & 20.243567625184863 \\
4 & 18.209840992660358 \\
5 & 16.497172446641568 \\
6 & 15.043246176657511 \\
7 & 13.799606631510756 \\
8 & 12.728249973986842 \\
9 & 11.799122653005792 \\
10 & 10.988269542104628 
\end{tabular}
\end{ruledtabular}
\end{table}
\section{Results and Discussions}\label{discuss}
The expression given for electron repulsion integrals in the Eq. (\ref{eq:16}) involves two different hyper$-$geometric functions. Calculation of the Eq. (\ref{eq:16}) through recurrence relationship method for the hyper$-$geometric functions due to nature of these special functions generates multiple hyper$-$geometric functions. This increases the complications during the computation of two electron integrals. A simple recurrence relationship which runs directly over the radial integrals (Eq. (\ref{eq:40})) and depending to $L$ is derived. It is available to be coded in Fortran programming language and no additional packages such as packages for computation of special functions, are needed. Additionally, all the integrals of $\mathfrak{R}^{L}_{n,n^{\prime}}$ are calculated simultaneously over $L$. The initial conditions are obtained by calculating only one hyper$-$geometric for $L=0$. The advantageous of the method proposed here, are particularly evident when performing calculations for heavy atoms.\\
The hyper$-$geometric functions are also represented in terms of radial integrals $\mathfrak{R}^{L}_{n,n^{\prime}}$. This allows to check the computation times for both hyper$-$geometric functions and electron repulsion integrals. In the \textcolor{red}{Figure \ref{fig:Hypercpu}} the CPU time for calculation of left and right$-$hand sides of the Eq. (\ref{eq:34}) is plotted. The time to obtain a list of results depending to $L$ is compared. Fist ten elements of these list is given in the Table \ref{tab:Hyperlist}. The Mathematica programming language is used for calculations. The list for the left$-$hand side of the Eq. (\ref{eq:34}) is obtained by using standard library of Mathematica for special functions and using the command \textbf{Table}. The list for the right$-$hand side is obtained by using the command \textbf{RecurrenceTable}. The figure shows that the method presented here, for computation of electron repulsion integrals is robust. It also shows that, even the standard Mathematica library for special functions is much slower than our method. Note that, according to Eq. (\ref{eq:8}) $L< \left\{n, n^{\prime}\right\}$. Since the CPU times for hyper$-$geometric functions are compared such restriction need not to be considered.

The Eq. (\ref{eq:13}) for two center case is expressed in prolate spheroidal coordinates. The resulting auxiliary functions have similar symmetry with right$-$hand sides of the Eqs. (\ref{eq:14}, \ref{eq:15}) but in two$-$dimensions. They are given as \cite{34_Bagci_2022} (See also references therein),
\begin{multline} \label{eq:58}
\left\lbrace \begin{array}{cc}
\mathcal{P}^{n_1,q}_{n_{2}n_{3}n_{4}}\left(p_{123} \right)
\\
\mathcal{Q}^{n_1,q}_{n_{2}n_{3}n_{4}}\left(p_{123} \right)
\end{array} \right\rbrace
\\
=\frac{p_{1}^{\sl n_{1}}}{\left({\sl n_{4}}-{\sl n_{1}} \right)_{\sl n_{1}}}
\int_{1}^{\infty}\int_{-1}^{1}{\left(\xi\nu \right)^{q}\left(\xi+\nu \right)^{\sl n_{2}}\left(\xi-\nu \right)^{\sl n_{3}}}\\ \times
\left\lbrace \begin{array}{cc}
P\left[{\sl n_{4}-n_{1}},p_{1}(\xi+\nu) \right]
\\
Q\left[{\sl n_{4}-n_{1}},p_{1}(\xi+\nu) \right]
\end{array} \right\rbrace
e^{p_{2}\xi-p_{3}\nu}d\xi d\nu,
\end{multline}
The Eqs. (\ref{eq:14}, \ref{eq:15}) are expressed in terms of Gauss's hyper$-$geometric functions. Finite series representations for the Eq. (\ref{eq:58}) as given in previous section for electron repulsion integrals may be possible but only if corresponding special functions are found. The relationships given in the Eqs. (\ref{eq:14}, \ref{eq:15}) are obtained by series representation of lower incomplete gamma functions,
\begin{align}\label{eq:59}
\gamma\left[a,x\right]
=x^{a}\Gamma\left(a\right)e^{-x}
\sum_{k=0}^{\infty}\dfrac{x^{k}}{\Gamma\left(a+k+1\right)},
\end{align}
and, the following integrals of elementary functions,
\begin{align}\label{eq:60}
\int_{0}^{\infty}x^{a}e^{-\zeta x}dx=\Gamma\left(a+1\right)\zeta^{-a-1}.
\end{align}
The expression for integrals involve the upper incomplete gamma functions (Eq. (\ref{eq:15})) is then, derived by using the recurrence relationships for hyper$-$geometric functions,
\begin{multline}\label{eq:61}
{_2}F_{1}\left[a,b,c;z\right]
=\dfrac{\Gamma\left(c\right)\Gamma\left(a+b-c\right)}{\Gamma\left(a\right)\Gamma\left(b\right)}\left(1-z\right)^{c-a-b}
\\
\times {_2}F_{1}\left[c-a,c-b,c-a-b+1;1-z\right]
\\
+\dfrac{\Gamma\left(c\right)\Gamma\left(c-a-b\right)}{\Gamma\left(c-a\right)\Gamma\left(c-b\right)}
{_2}F_{1}\left[a,b,a+b-c+1;1-z\right].
\end{multline}
Implementing a similar routine to Eq. (\ref{eq:58}) requires convergent series representation for power series such as, $\left(x+y\right)^{a}$, $a \in \mathbb{R}$, for both $\vert x/y \vert <1$ and $\vert x/y \vert >1$. The following binomial expansion satisfies this requirements,
\begin{multline}\label{eq:62}
\left(x \pm y\right)^{\lfloor a \rfloor+\epsilon}
=\left(\pm 1\right)^{\delta^{a}_{\epsilon}}
\lim_{N \to \infty}
\sum_{k=0}^{\lfloor a \rfloor+N}
\left(\pm 1\right)^{k}
\\
\times \dfrac{\Gamma\left(\lfloor a \rfloor+\epsilon+1\right)}{\Gamma\left(k-\delta^{\epsilon}_{N}+\epsilon+1\right)\Gamma\left(\lfloor a \rfloor-k+\delta_{N}+1\right)}
\\
\times x^{k-\delta^{\epsilon}_{N}+\epsilon}a^{\lfloor a \rfloor+\delta^{\epsilon}_{N}-k},
\end{multline}
\begin{align}\label{eq:55}
\delta^{\epsilon}_{N}=\left\{
\begin{array}{ll}
N \hspace{3mm} \vert x/y \vert >1
\\
\epsilon  \hspace{5mm} \vert x/y \vert <1
\end{array}
\right.
\end{align}
$\delta^{a}_{\epsilon}=\epsilon$ for $x-y<0$ and $\delta^{a}_{\epsilon}=\lfloor a \rfloor$ for $x-y>0$.\\
The problem about the conditional existence of binomial expansion can be eliminated by writing the Eq. (\ref{eq:58}) in a two$-$range form depending on domain for $\nu$ as $-1\leq \nu \leq 0$ and $0\leq \nu \leq 1$ and using the following expressions,
\begin{align}\label{eq:56}
\left(\xi+\nu\right)^{2}
=\left(\xi-\nu\right)^{2}+4\left(\xi \nu\right),
\end{align} 
for $-1\leq \nu \leq 0$ and,
\begin{align}\label{eq:57}
\left(\xi-\nu\right)^{2}
=\left(\xi+\nu\right)^{2}-4\left(\xi \nu\right),
\end{align}
for $0\leq \nu \leq 1$.\\
As corresponding special functions for the Eq. (\ref{eq:58}) we expect to have another special functions such as Appell functions but we remain this task for next paper of our series.

\section*{Acknowledgement}
One of the authors A.B. acknowledges funding for a postdoctoral research fellowship from CONICET-IMIT Instituto de Modelado e Innovaci{\'o}n Tecnol{\'o}gica (Consejo Nacional de Investigaciones Científicas y T{\'e}cnicas - Universidad Nacional del Nordeste)
\pagebreak
\appendix
\section{Recurrence relationships for $l_{nn^{\prime}}^{L}$}\label{recl}
Recurrence relationships for the finite summation formula $l_{nn^{\prime}}^{L}$ that arise in $m^{L}_{nn^{\prime}}$, are given for seek of completeness and self$-$compatibility of the method.\\
We start by representing the $l_{nn^{\prime}}^{L}$ in terms of hyper$-$geometric functions:
\begin{multline}\label{eq:a1}
{^{2}}f^{L}_{nn^{\prime}}\left(\zeta, \zeta^{\prime}\right)
=\dfrac{\pi csc\left[\left(-n^{\prime}+L\right)\pi\right]}{\Gamma\left(-n^{\prime}+L+1\right)}
\dfrac{\left(n+L+1\right)}{\Gamma\left(n^{\prime}-L+1\right)}
\\
\times \left(-\dfrac{\zeta^{\prime}}{\zeta}\right)^{2L+1}
\left(\dfrac{\zeta+\zeta^{\prime}}{\zeta^{\prime}}\right),
\end{multline}
\begin{multline}\label{eq:a2}
\sum_{k=1}^{2L+1}\dfrac{\left(n-L+k\right)_{2L+1-k}}{\left(-n^{\prime}-L-1+k\right)_{2L+1-k}}\left(-\dfrac{\zeta}{\zeta^{\prime}}\right)^{k-1}
\\
=\Gamma\left(n+L+1\right)\dfrac{\Gamma\left(-n^{\prime}-L\right)}{\Gamma\left(-n^{\prime}+L\right)}\dfrac{1}{\Gamma\left(n-L+1\right)}
\\
\times {_{2}}F_{1}\left[1,-n^{\prime}-L,n-L+1; \frac{\zeta}{\zeta^{\prime}}\right]
\\
-\Gamma\left(n+L+1\right)\dfrac{\left(-n^{\prime}+L\right)}{\Gamma\left(n+L+2\right)}\left(\dfrac{\zeta}{\zeta^{\prime}}\right)^{2L+1}
\\
\times {_{2}}F_{1}\left[1,-n^{\prime}+L+1,n+L+2; \frac{\zeta}{\zeta^{\prime}}\right].
\end{multline}
The Eq. (\ref{eq:23}) and the following recurrence relationship are now available to be used,
\begin{multline}\label{eq:a3}
{_{2}}F_{1}\left[1,b,c;z\right]
=\dfrac{\left(1-b-m\right)_{m}}{\left(c-b-m\right)_{m}}\left(1-z\right)^{m}
{_{2}}F_{1}\left[1,b+m,c;z\right]
\\
-
\dfrac{\left(1-b-m\right)_{m}}{\left(c-b-m\right)_{m}}\left(1-z\right)^{m}
\left(\dfrac{c-1}{b+m-1}\right)
\\
\times 
\sum_{k=0}^{m-1}\dfrac{\left(c-b-m\right)_{k}}{\left(2-b-m\right)_{k}}
\left(1-z\right)^{-k-1}.
\end{multline}
The following expressions are obtained as results,\\
over $n$,
\begin{multline}\label{eq:a4}
{^{1}}l^{L}_{n+1,n^{\prime}}\left(\zeta, \zeta^{\prime}\right)
=\dfrac{\left(n+L+2\right)}{\left(n+n^{\prime}+1\right)}\left(\dfrac{\zeta^{\prime}}{\zeta}\right)\left(\dfrac{\zeta+\zeta^{\prime}}{\zeta^{\prime}}\right)
{^{1}}l^{L}_{n,n^{\prime}}\left(\zeta, \zeta^{\prime}\right)
\\
+\dfrac{\left(n+L+2\right)}{\left(n+L+1\right)}
\dfrac{1}{\left(n+n^{\prime}+1\right)}
{^{2}}f^{L}_{nn^{\prime}}\left(\zeta, \zeta^{\prime}\right)
\\
\times \bigg\{
\dfrac{\left(n+L+2\right)}{\left(n-L+1\right)}
\dfrac{\left(-n^{\prime-L}\right)}{\left(-n^{\prime}+L\right)}
\left(-\dfrac{\zeta^{\prime}}{\zeta}\right)
\\
-\left(n^{\prime}+L\right)\left(-\dfrac{\zeta}{\zeta^{\prime}}\right)^{2L}
\bigg\}
\end{multline}
over $n^{\prime}$,
\begin{multline}\label{eq:a5}
{^{1}}l^{L}_{n,n^{\prime}+1}\left(\zeta, \zeta^{\prime}\right)
\dfrac{\left(n^{\prime}-L+1\right)}{\left(n+n^{\prime}+1\right)}
\left(\dfrac{\zeta+\zeta^{\prime}}{\zeta^{\prime}}\right)
{^{1}}l^{L}_{n,n^{\prime}}\left(\zeta, \zeta^{\prime}\right)
\\
\dfrac{\left(n^{\prime}-L\right)}{\left(n^{\prime}-L+1\right)}\dfrac{1}{\left(n+n^{\prime}+1\right)}
\\
\times
\bigg\{
\dfrac{\Gamma\left(n+L+1\right)}{\Gamma\left(n-L\right)}
\dfrac{\Gamma\left(-n^{\prime-L-1}\right)}{\Gamma\left(-n^{\prime}+L-1\right)}
\\
-\left(-n^{\prime}+L-1\right)\left(\dfrac{\zeta}{\zeta^{\prime}}\right)^{2L+1}
\bigg\}
\end{multline}
over $L$,
\begin{multline}
{^{1}}l^{L+1}_{n,n^{\prime}}\left(\zeta, \zeta^{\prime}\right)
=
\dfrac{\left(-n^{\prime}+L\right)}{\left(-n^{\prime}+L+2\right)}
\left(\dfrac{n+L+2}{n+L+1}\right)^{2}
\\
\dfrac{\left(-n^{\prime}+L\right)}{\left(-n^{\prime}+L+1\right)}
{^{1}}l^{L}_{n,n^{\prime}}\left(\zeta, \zeta^{\prime}\right)
\\
+\dfrac{\left(-n^{\prime}+L\right)}{\left(-n^{\prime}+L+2\right)}
\left(\dfrac{n+L+2}{n+L+1}\right){^{2}}f^{L}_{nn^{\prime}}\left(\zeta, \zeta^{\prime}\right)
\\
\bigg\{
\left(-\dfrac{\zeta}{\zeta^{\prime}}\right)^{2L}
+\dfrac{\left(n+L+1\right)}{\left(-n^{\prime}+L\right)}
\left(-\dfrac{\zeta}{\zeta^{\prime}}\right)^{2L-1}
\\
+\left(\dfrac{n+L+2}{n+L+1}\right)
\dfrac{\left(-n^{\prime}+L\right)}{\left(-n^{\prime}+L+1\right)}
\dfrac{\left(n-L\right)}{\left(n+L+1\right)}
\bigg\}.
\end{multline}

\end{document}